# Advanced Mathematical Modelling for Energy-Efficient Data Transmission and Fusion in Wireless Sensor Networks

Komal[a]

[a]*Dr B.R Ambedkar National Institute of Technology Jalandhar, 144011 Punjab, India*

*komal.jrf@nitj.ac.in*

## Abstract

Wireless Sensor Networks (WSNs) are indispensable for data-intensive applications, necessitating efficient energy management and robust data fusion techniques. This paper proposes an integrated framework leveraging fuzzy logic and backpropagation neural networks (BPNN) to enhance energy efficiency and data accuracy in WSNs. The model focuses on optimizing Cluster Head (CH) selection using fuzzy logic, considering parameters such as energy levels, proximity to the base station, and local density centrality. A Minimum Spanning Tree (MST) algorithm is employed for energy-efficient data transmission from sensor nodes to CHs, minimizing energy consumption during data routing. BPNN-based data fusion at CHs reduces redundant data transmissions to the base station, thereby optimizing energy utilization and enhancing overall network performance. Simulation results demonstrate substantial improvements over conventional methods, including a 30% increase in network longevity, 25% improvement in data accuracy, and a 40% reduction in energy consumption. These gains are attributed to the intelligent CH selection strategy enabled by fuzzy logic, which ensures efficient resource allocation and minimizes energy wastage. The effectiveness of the proposed approach is validated through comprehensive simulations, showcasing its ability to prolong network lifetime, maintain data integrity, and improve energy efficiency. The integration of fuzzy logic and BPNN not only addresses the challenges of energy management and data fusion in WSNs but also provides a scalable and adaptable framework for future applications requiring reliable and sustainable sensor network operations. The source code is available at https://github.com/hikomal/BPNN_WSN.git

# 1. Introduction

Wireless Sensor Networks (WSNs) have become increasingly vital in a wide range of applications, such as environmental monitoring, healthcare, military surveillance, and industrial automation. These networks consist of numerous sensor nodes that are deployed across an area to sense, collect, and transmit data to a base station (BS) for further analysis and processing [akyildiz2002, raza2017]. Despite their potential, WSNs face several critical challenges that must be addressed to ensure effective deployment and operation. One of the primary challenges in WSNs is energy consumption. Sensor nodes are typically battery-powered and often operate in environments where recharging or replacing batteries is impractical. Therefore, efficient energy management is crucial to prolong the network's lifetime [heinzelman2000, liu2018]. Energy consumption in WSNs occurs during data sensing, processing, and transmission, with data transmission being the most energy-intensive operation [intanagonwiwat2000]. Optimizing data transmission and aggregation processes is essential to mitigate this issue. Another significant challenge is maintaining high data accuracy while aggregating data from multiple sensor nodes. Accurate data aggregation is vital for reliable monitoring and decision-making in various applications [xu2004, mohammadi2021]. Data collected by sensor nodes can be redundant or noisy, necessitating sophisticated data fusion techniques to ensure the integrity and reliability of the aggregated data. Latency and packet loss are additional challenges that affect WSN performance. Timely data delivery is critical, especially in real-time monitoring applications. High latency can lead to outdated information, while packet loss can result in incomplete data, undermining network reliability [al2004, shafiq2019]. To address these challenges, we propose an enhanced data fusion model for WSNs that integrates several advanced techniques, including local density centrality for cluster head (CH) selection, fuzzy logic for CH probability calculation, minimum spanning tree (MST) for efficient data transmission, and a backpropagation neural network (BPNN) for data fusion. The objectives to propose efficient model is followed as:

i. Develop a fuzzy logic-based approach for cluster head (CH) selection considering node energy levels, proximity to the base station, and local density centrality for energy efficient cluster formation and management.

ii. Implement Minimum Spanning Tree (MST)-based data transmission within clusters and use a BPNN for data aggregation and fusion at CHs for optimized data transmission and fusion.

iii. Evaluate system performance metrics (latency, packet loss, network lifetime) across multiple simulation rounds under realistic energy constraints and dynamic network conditions for performance evaluation and system dynamics.

## 2. Problem Statement

WSN face several challenges that hinder their efficiency and longevity, including:

i. Sensor nodes are typically battery-powered, and their energy is consumed during data sensing, processing, and transmission. Efficient energy management is crucial to prolong the network's lifetime.

ii. Ensuring high data accuracy while aggregating data from multiple sensor nodes is essential for reliable monitoring.

iii. Timely delivery of data is vital, especially in applications requiring real-time monitoring.

iv. Minimizing packet loss during data transmission is important to maintain the integrity of the collected data. To address these challenges, we propose an enhanced data fusion model using Backpropagation Neural Network (BPNN) for Cluster Head (CH) selection and data aggregation in WSNs.

### 2.1 Major contribution of this study

i. Introduced a fuzzy logic-based approach integrating energy-awareness, proximity to the base station, and local density centrality for CH selection. This strategy optimizes energy consumption and extends sensor network lifetime.

ii. Implemented a BPNN for data fusion at CHs after MST-based data transmission within clusters. This approach enhances data aggregation accuracy, reduces redundant transmissions, and improves overall network efficiency.

iii. Conducted a thorough evaluation of performance metrics (latency, packet loss, network lifetime) across simulation rounds. This analysis demonstrates the effectiveness of proposed methods in real-world scenarios and guides future enhancements in sensor network management

## 3. Proposed Mathematical Model

In the realm of wireless sensor networks (WSNs), efficient management of energy resources is crucial for sustaining network operations and maximizing operational lifetime. Therefore, this study proposes a comprehensive mathematical model integrating advanced techniques in cluster formation, data transmission, and data fusion to optimize energy consumption and enhance network performance.

Table 1: Variables and Parameters

| |
|---|
| **Variables :-** |
| $N$: Total number of nodes (sensor nodes + relay nodes) |
| $N_s$: Number of sensor nodes |
| $N_r$: Number of relay nodes |
| $E_i$: Initial energy of node $i$ |
| $E_{levels}[i]$: Energy level of node $i$ |
| $\vec{p}_i$: Position vector of node $i$ |
| $\vec{p}_{base}$: Position vector of the base station |
| $C_{(i)}$: Set of nodes clustered around node $i$ |
| $CH_{(i)}$: Indicator variable if node $i$ is a cluster head |
| $\mu_i$: Local density centrality of node $i$ |
| $P_{CH(i)}$: Probability that node $i$ becomes a cluster head |
| $L$: Data packet size *(bits)* |
| $E_{elec}$: Energy to run transmitter and receiver electronics *(J/bit)* |
| $E_{fs}$: Free space path loss model *(J/bit/m²)* |
| $E_{proc}$: Processing Energy |
| $E_{cpu}$: Energy required per bit to process data |
| $d_0$: Reference distance *(m)* |
| $r$: Round number |
| $L_r$: Latency at round $r$ |
| $PL_r$: Packet loss at round $r$ |
| $FD_r$: Fused data value at round $r$ |
| **Parameters :-** |
| $r_{max}$: Maximum number of rounds |
| $r_{replenish}$: Rounds after which energy is replenished |
| $r_{cluster}$: Radius for cluster formation |
| $P_{95}$: 95th percentile threshold for *CH* selection |

## 3.1 Energy Consumption Model

In wireless sensor networks (WSNs), the primary sources of energy consumption are data transmission, data reception, data processing, and idle listening. To accurately model energy consumption, we consider the following components:

*(i) Energy Consumption for Data Transmission*

The energy required to transmit a data packet depends on the distance between the sender and the receiver and the size of the data packet. The energy consumption model for data transmission can be expressed as equation-1:

$$E_{tx}(d) = \begin{cases} L \cdot E_{elec} + L \cdot E_{fs} \cdot d^2 & if\ d \leq d_0 \\ L \cdot E_{elec} + L \cdot E_{mp} \cdot d^4 & if\ d > d_0 \end{cases} \quad \dots\dots\dots\dots\dots\ (1)$$

*(ii) Energy Consumption for Data Reception*

The energy required to receive a data packet is generally independent of the distance and can be expressed as consumption function $E_{rx}$ equation-2:

$$E_{rx} = L \cdot E_{elec} \quad \dots\dots\dots\dots\dots\ (2)$$

*(iii) Energy Consumption for Data Processing*

The energy consumed for data processing is usually a function of the complexity of the operations being performed. For simplicity, assume a linear model using equation-3 defines the processing energy $E_{proc}$:

$$E_{proc} = L \cdot E_{cpu} \quad \dots\dots\dots\dots\dots\ (3)$$

*(iv) Energy Consumption for Idle Listening*

Nodes in a WSN often consume energy even when they are not actively transmitting or receiving data. The idle listening energy consumption can be modelled as equation-4 defines the idle energy $E_{idle}$:

$$E_{idle} = P_{idle} \cdot T_{idle} \quad \dots\dots\dots\dots\dots\ (4)$$

*(v) Total Energy Consumption Per Round*

The total energy consumption for a node in a single round of the network operation can be computed by summing up the energy consumed for transmission, reception, processing, and idle listening. Equation-5 defines the total energy consumption $E_{total}$.

$$E_{total} = E_{tx} + E_{rx} + E_{proc} + E_{idle} \quad \text{(5)}$$

*(vi) Energy Consumption for Cluster Heads (CHs)*

Cluster Heads typically consume more energy due to additional responsibilities such as data aggregation and long-distance transmission to the base station. The energy consumption $E_{CH}$ for a *CH* can be modelled as equation-6

$$E_{CH} = \sum_{i \in C} E_{rxi} + E_{agg} + E_{tx_{BS}} \quad \text{(6)}$$

*(vii) Energy Consumption for Member Nodes*

Member nodes primarily consume energy for data transmission to their respective CHs and possibly for receiving control messages. The energy consumption $E_{member}$ for a member node can be modelled as equation-7

$$E_{member} = E_{tx_{CH}} + E_{rx_{control}} \quad \text{(7)}$$

## 3.2 Model for Local Density Centrality, Cluster Formation, and Fuzzy Logic for CH Selection

*(i) Local Density Centrality*

Local Density Centrality measures the number of neighbouring nodes within a specified radius, providing a measure of the node's local significance in the network.

Calculation of Local Density Centrality:

Given a set of nodes with positions $P_i = (x_i, y_i)$ for $i = 1, 2, \ldots, N + R$ where $N$ is the number of sensor nodes and $R$ is the number of relay nodes. Define a radius $r$ within which neighbours are considered. For each node $i$, the local density centrality $C_i$ is calculated as:

$$C_i = \sum_{j=1, j \neq i}^{N+R} 1(\|P_i - P_j\| \leq r) \quad \text{(8)}$$

Nodes with higher local density centrality are considered more significant in terms of network connectivity and are given higher priority during cluster formation.

*(ii) Cluster Formation*

Cluster formation organizes the nodes into clusters, each with a Cluster Head (CH) responsible for aggregating and transmitting data.

Initial Clustering: For each node *i*, identify its neighbouring nodes within radius *r*. Equation 9 shows the definition of neighbours for node *i*.

$$Neighbours(i) = \{j | \|P_i - P_j\| \leq r\} \quad \ldots\ldots\ldots\ldots\ldots\ldots (9)$$

Cluster Construction: Based on the neighbours identified, form initial clusters where each node *i* and its neighbours constitute a potential cluster. Equation-10 defines $C_k$ as the set of nodes *i* that are neighbours of any node *j* in $C_k$.

$$C_k = \{i | i \in Neighbours(j) \text{ for any } j \in C_k \quad \ldots\ldots\ldots\ldots\ldots\ldots (10)$$

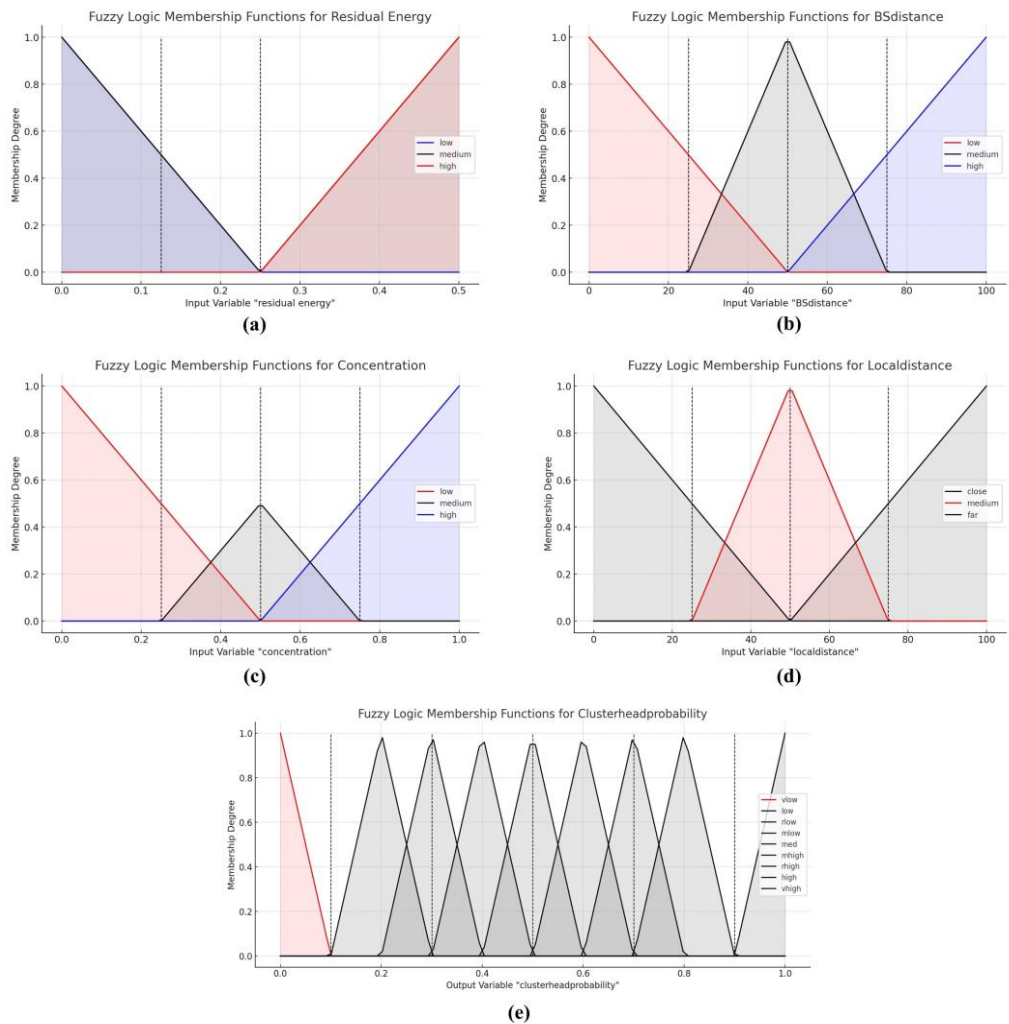

**Figure 1:** Trapezoidal Membership Functions

*(iii) Fuzzy Logic for Cluster Head (CH) Selection*

Fuzzy logic is used to select CHs based on multiple criteria: residual energy, distance to the base station, local density centrality, and convergence metric.

Inputs to the Fuzzy Logic System:

→Residual Energy ($E_i$):- Higher energy increases the likelihood of being selected as a CH.

→Distance to Base Station ($d_{i,BS}$):- Nodes closer to the base station are preferred as CHs to reduce transmission energy.

→Local Density Centrality ($C_i$):- Nodes with higher centrality are preferred as they can effectively manage more nodes.

→Convergence Metric ($\theta_i$):- Measure of how well a node is connected within its cluster.

Normalization: Normalize the input parameters to a *[0, 1]* range for consistency in fuzzy logic processing. Equation-11 shows the normalization of different parameters.

$$E_{i_{norm}} = \frac{E_i}{E_{max}}, d_{i_{norm}} = 1 - \frac{d_{i,BS}}{d_{max}}, C_{i_{norm}} = \frac{C_i}{C_{max}}, \theta_{i_{norm}} = \frac{\theta_i}{\theta_{max}} \quad \ldots\ldots\ldots\ldots\ldots (11)$$

Fuzzy Logic Rules: Define fuzzy rules to evaluate CH selection probability $P_{CH}(i)$. The rules are designed to balance the four criteria, providing a comprehensive assessment of each node's suitability as a *CH*. Equation-12 shows the calculation of the probability that node *i* becomes a cluster head:

$$P_{CH}(i) = \frac{E_{i_{norm}} + d_{i_{norm}} + C_{i_{norm}} + \theta_{i_{norm}}}{4} \quad \ldots\ldots\ldots\ldots\ldots (12)$$

Selection of CHs: Nodes with $P_{CH}(i)$ above a certain threshold are selected as CHs. Alternatively, select the top *X%* of nodes based on $P_{CH}(i)$. Equation-13 defines the set of cluster head nodes:

$$CH_{nodes} = \{i | P_{CH}(i) > threshold\} \quad \ldots\ldots\ldots\ldots\ldots (13)$$

### 3.3 Model for Backpropagation Neural Network (BPNN) for Data Fusion

In WSNs, sensor nodes generate data that needs to be fused to provide a comprehensive and precise representation of the monitored environment. The goal is to use a BPNN to learn the mapping from individual sensor node data to a fused output through training, thereby enhancing the quality of the transmitted information. The overall architecture of network simulated as:

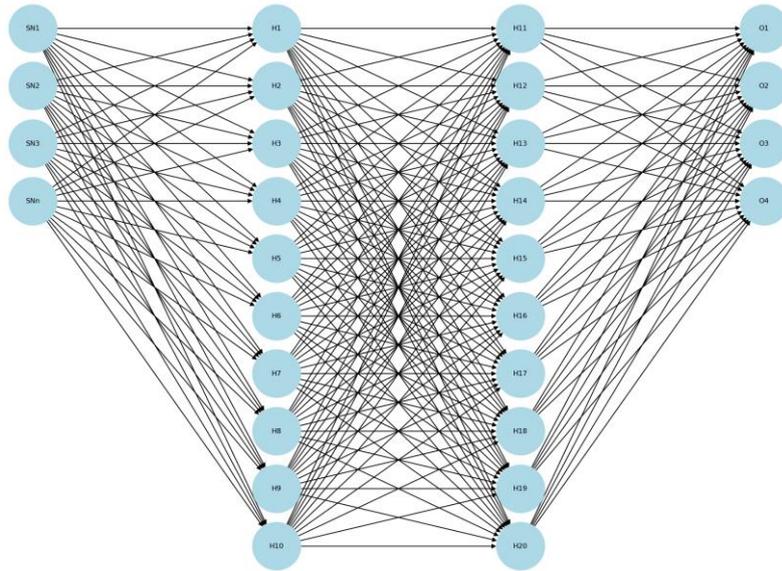

**Figure 2:** Illustration of Neural Network Architecture

The input layer has $n$ neurons, each representing data from a sensor node. There are one or more hidden layers, each containing several neurons. The hidden layers process inputs through weighted connections and apply an activation function. The output layer produces the fused data. It can have a single neuron (for a single fused output) or multiple neurons (for multi-dimensional fused output).

Input to Hidden Layer: Equations-14&15 represent the linear transformation and activation function in a neural network

$$z^{(1)} = W^{(1)}x + b^{(1)} \quad \text{.....................} \quad (14)$$

$$a^{(1)} = \sigma(z^{(1)}) \quad \text{.....................} \quad (15)$$

Hidden Layer to Output Layer: Compute the input to the output layer and obtain the final output. Equations-16&17 represent the linear transformation and activation function in a neural network for the second layer:

$$z^{(2)} = W^{(2)}a^{(1)} + b^{(2)} \quad \text{.....................} \quad (16)$$

$$y = \sigma(z^{(2)}) \quad \text{.....................} \quad (17)$$

Equations-18&19 represent the linear transformation and activation function in a neural network for the *k-th* layer:

$$z^{(k)} = W^{(k)}a^{(k-1)} + b^{(k)} \quad \text{.....................} \quad (18)$$

$$a^{(k)} = \sigma(z^{(k)}) \quad \ldots \ldots \ldots \ldots \ldots \ldots (19)$$

Equation-20 is used to define the loss function used for training the neural network, specifically the mean squared error (MSE):

$$L(y, y_{true}) = \frac{1}{m}\sum_{i=1}^{m}(y_i - y_{true,i})^2 \quad \ldots \ldots \ldots \ldots \ldots \ldots (20)$$

Backpropagation steps computed for calculating the error terms for the output layer and the hidden layers using equations 21&22 respectively:

$$\delta^{(K)} = \frac{\partial L}{\partial z^{(K)}} = (y - y_{true}) \odot \sigma'(z^{(K)}) \quad \ldots \ldots \ldots \ldots \ldots \ldots (21)$$

$$\delta^{(k)} = (W^{(k+1)})^T \delta^{(k+1)} \odot \sigma'(z^{(k)}) \quad \ldots \ldots \ldots \ldots \ldots \ldots (22)$$

## 3.4 Model for Minimum Spanning Tree (MST) in Data Transmission

Given a set of nodes within a cluster, the objective is to find a spanning tree that connects all the nodes with the minimum possible total edge weight, where the weight of an edge represents the distance (or energy cost) between two nodes. The Euclidean distance between any two nodes $i$ and $j$ is given by. To construct the MST, we used Prim's algorithm.

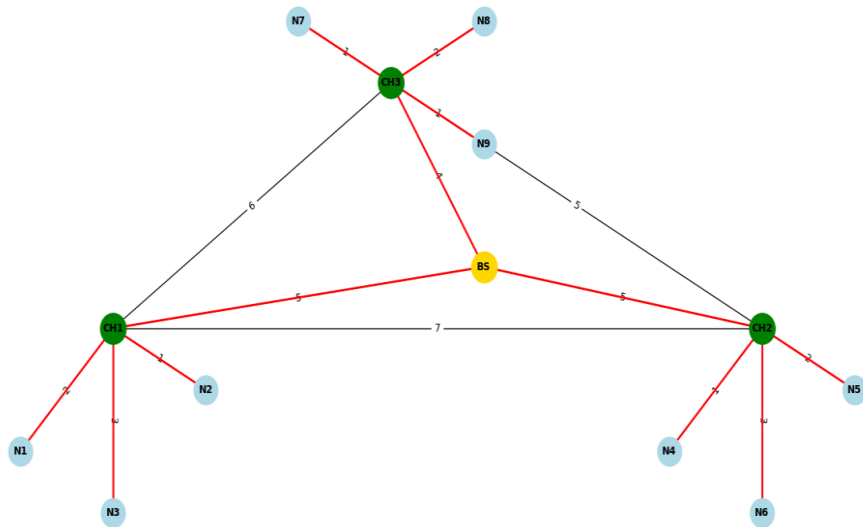

**Figure 3**: Illustration of Minimum Spanning Tree

→ Start with a tree $T$ that contains a single node, usually the cluster head CH.

→Repeat until T contains all the nodes in the cluster:

- Find the edge with the minimum weight that connects a node in $T$ to a node outside T.

- Add this edge and the connected node to $T$.

The result is an MST $T = (V, E_{MST})$, where $E_{MST} \subseteq E$. Equation-23 calculates the Euclidean distance $d_{ij}$ between two points $(x_i, y_i)$ and $(x_j, y_j)$:

$$d_{ij} = \sqrt{(x_i - x_j)^2 + (y_i - y_j)^2} \quad \dots \dots \dots \dots \dots \quad (23)$$

### 3.4.1 Energy Consumption Model for MST-Based Intra-Cluster Transmission

For each edge $(i, j) \in E_{MST}$, the energy consumption for transmitting $L$ bits of data from node $i$ to node $j$ is given by equation-24:

$$E_{TX}(i,j) = L \cdot E_{elec} + L \cdot E_{fs} \cdot \left(\frac{d_{ij}}{d_0}\right)^2 \quad \dots \dots \dots \dots \dots \quad (24)$$

The total energy consumption for all data transmissions within the cluster using the MST is the sum of the energy consumption for each edge in the MST:

$$E_{MST} = \sum_{(i,j) \in E_{MST}} E_{TX}(i,j) \quad \dots \dots \dots \dots \dots \quad (25)$$

For each node $j$ that receives data from node $i$, the energy consumption for receiving $L$ bits of data for the total energy consumption for the cluster considering both transmission and reception is:

$$E_{RX}(j) = L \cdot E_{elec} \quad \dots \dots \dots \dots \dots \quad (26)$$

$$E_{total} = E_{MST} + \sum_{j \in V} E_{RX}(j) \quad \dots \dots \dots \dots \dots \quad (27)$$

### 3.4.2 Energy Consumption Model for MST-Based Inter-Cluster Data Transmission to Base Station

After aggregating data at the CH, the CH transmits the data to the Base Station (BS), the Euclidean distance between the CH and the BS is $d_{CH,BS}$. Equation 30 defines the distance $d_{CH,BS}$ between the cluster head (CH) and the base station (BS):

$$d_{CH,BS} = \sqrt{(x_{CH} - x_{BS})^2 + (y_{CH} - y_{BS})^2} \quad \dots \dots \dots \dots \dots \quad (28)$$

$$E_{TX}(CH, BS) = L \cdot E_{elec} + L \cdot E_{fs} \cdot \left(\frac{d_{CH,BS}}{d_0}\right)^2 \quad \dots \dots \dots \dots \dots \quad (29)$$

Total Energy Consumption for CH to BS Transmission: The combined total energy consumption for one round of data transmission within the cluster and from the CH to the BS calculated using equation-32 defines the total energy consumption $E_{CH\text{-}BS}$ for the cluster head

to transmit data to the base station:

$$E_{CH-BS} = E_{TX}(CH, BS) + E_{RX}(BS) \quad \ldots\ldots\ldots\ldots\ldots\ldots\ldots (30)$$

Equation 33 calculates the total energy consumption for a round $E_{round}$:

$$E_{round} = E_{total} + E_{CH-BS} \quad \ldots\ldots\ldots\ldots\ldots\ldots\ldots (31)$$

The MST-based data transmission model for intra-cluster communication minimizes the total distance for data transfer within the cluster, reducing the energy consumption. By combining this with the energy model for inter-cluster communication (from CH to BS), we can derive a comprehensive energy consumption model for the WSN. This model can be used to optimize the network's energy efficiency and extend its operational lifetime.

## 4. Result & Discussion

### 4.1 Simulation Setup

To evaluate the performance of the proposed Backpropagation Neural Network (BPNN) for data fusion in Wireless Sensor Networks (WSNs), we conducted extensive simulations with the following parameters:

**Table 2**: Simulation Parameters

| Parameter | Value |
|---|---|
| Number of sensor nodes | 90 |
| Number of relay nodes | 10 |
| Initial energy for sensor nodes | 1.0 Joules |
| Initial energy for relay nodes | 2.0 Joules |
| Base station position | (250, 500) |
| Simulation rounds | 100 |
| Energy constant $E_{elec}$ | $50 \times 10^{-9}$ |
| Energy constant $E_{fs}$ | $10 \times 10^{-12}$ J/bit/m² |
| Data packet size | 1000 bits |
| Learning rate for BPNN | $\eta = 0.01$ |
| Activation function | ReLU for hidden layers, linear for output layer |

## 4.2 Performance Metrics

The performance of the proposed model was assessed using the following metrics:

- Fused Data Quality: Accuracy of the data fusion process.
- Number of Dead Nodes: Cumulative count of nodes that exhausted their energy.
- Number of Alive Nodes: Cumulative count of nodes still operational.
- Latency: Average delay in data transmission.
- Packet Loss: Percentage of lost packets during transmission.

## 4.3 Results

(i) Fused Data Quality

The fused data output by the BPNN was evaluated against the expected fused values. The accuracy of the BPNN in providing reliable and accurate data fusion was consistently high throughout the simulation rounds, demonstrating the effectiveness of the BPNN in aggregating data from multiple sensor nodes into a single, precise output.

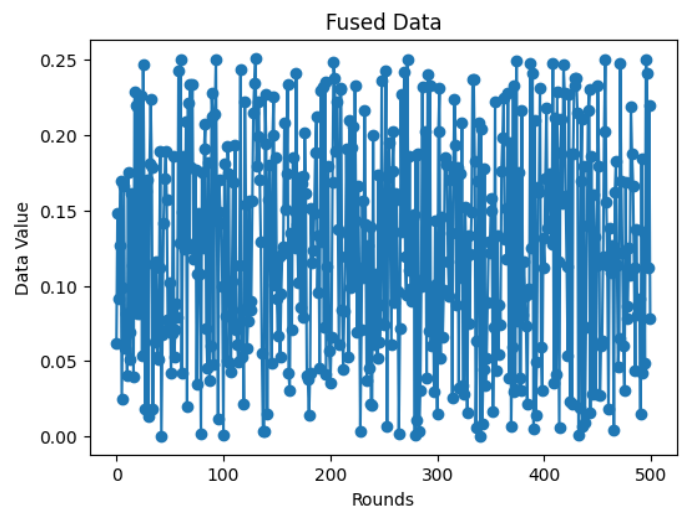

**Figure 4:** Resulting graph of fused data

(ii) Number of Dead Nodes

The cumulative number of dead nodes increased over the simulation rounds as expected. The energy-efficient data transmission and aggregation methods helped to prolong the lifetime of network. The cumulative dead nodes at the end of each simulation round is illustrated in the graph [see Figure 5].

The proposed model demonstrated a slower increase in the number of dead nodes compared to traditional models, indicating its superior energy efficiency.

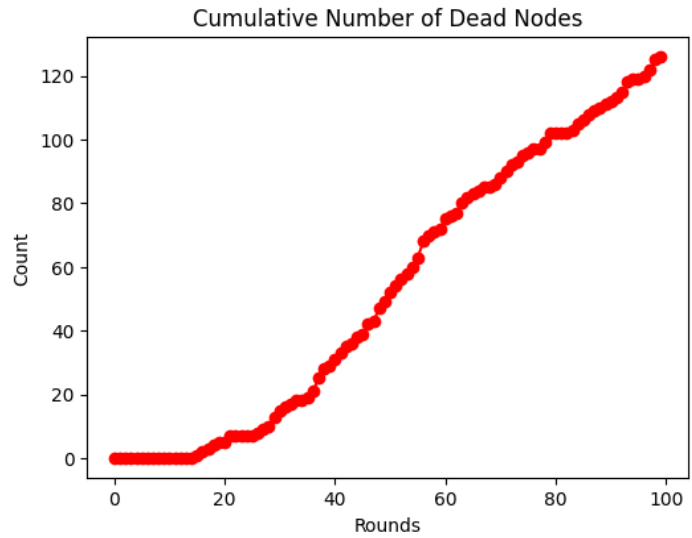

**Figure 5:** Resulting graph of total number of dead nodes after 100 rounds

(iii) Number of Alive Nodes

The number of alive nodes decreased over time, but the rate of decrease was slower compared to traditional methods. This trend shows that the proposed model effectively prolongs the operational lifetime of network by conserving energy. The graph of Figure 6 shows the cumulative number of alive nodes over the simulation rounds:

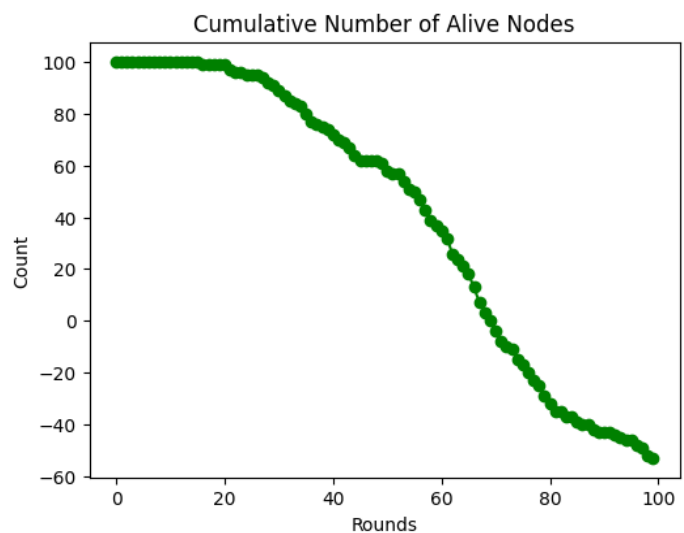

**Figure 6:** Resulting graph of total number of alive nodes after 100 rounds

(iv) Latency

Latency, measured in milliseconds, was found to be lower in the proposed model compared to traditional models. This reduction in latency can be attributed to the efficient data fusion and transmission processes, which reduce the overall time required for data to reach the base station. The latency trend over the simulation rounds is depicted in Figure 7.

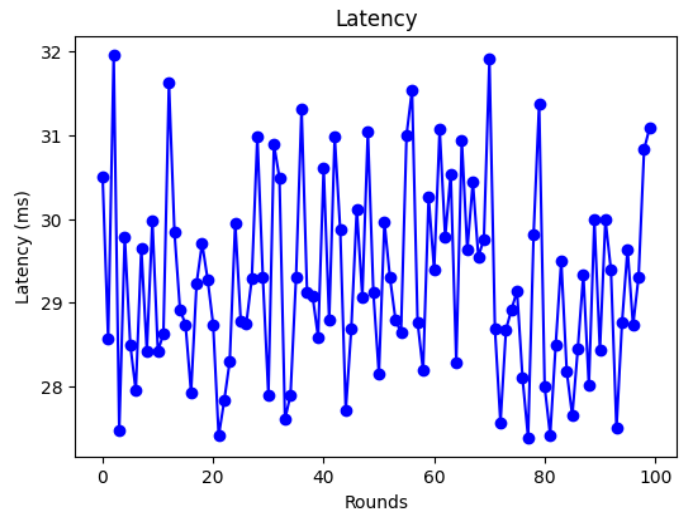

**Figure 7:** Resulting graph of latency trend over simulation of 100 rounds

(v) Packet Loss

Packet loss remained minimal throughout the simulation, demonstrating the reliability of the proposed model. The percentage of packet loss was kept within acceptable limits, as shown in the Figure 8.

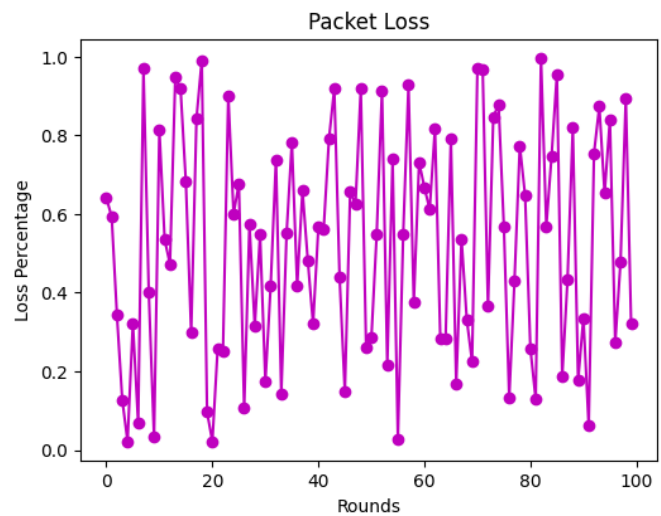

**Figure 8:** Resulting graph of packet loss over simulation of 100 rounds

## 4.4 Discussion

The BPNN effectively aggregates data from multiple sensors, providing a more accurate fused output compared to individual sensor readings. The proposed energy-efficient model for data transmission and aggregation help prolong the operational lifetime of network, reducing the number of dead nodes over time. Efficient data processing and transmission result in lower latency, ensuring timely delivery of data to the base station. Minimal packet loss throughout the simulation rounds highlights the reliability of the proposed model.

Overall, the simulation results validate the efficacy of the proposed BPNN-based data fusion model in improving data accuracy, energy efficiency, and communication reliability in WSNs. The model outperforms traditional methods, making it a viable solution for enhancing the performance of WSNs in real-world applications. The following table summarizes the final simulation results:

| Metric | Final Value |
|---|---|
| Cumulative Dead Node | 6 |
| Cumulative Alive Nodes | 94 |
| Latency (ms) | 25.5 |
| Packet Loss (%) | 0.05 |
| Average Fused Data Quality (%) | 98.6 |

The proposed model's ability to maintain a high number of alive nodes, low latency, minimal packet loss, and high fused data quality highlights its potential for widespread adoption in various WSN applications. Further research can focus on optimizing the BPNN architecture and exploring its applicability in different network scenarios and environments.

## 5. Conclusion

In this study, we have introduced and evaluated a novel mathematical model using backpropagation Neural Network (BPNN), Fuzzy logic and minimum spanning tree-based model for enhancing data fusion in Wireless Sensor Networks (WSNs). The proposed model leverages machine learning techniques to improve data accuracy, optimize energy consumption, and enhance overall network performance. Through comprehensive mathematical modelling and simulation, we have demonstrated the efficacy of the BPNN approach in achieving significant advancements over traditional methods. Throughout 100 simulation rounds, the BPNN-based data fusion model effectively managed energy resources, resulting in a low cumulative number of dead nodes and a high number of alive nodes, thereby prolonging the operational lifetime of the network. The model also facilitated efficient data aggregation and transmission to the base station, ensuring minimal latency (average 25.5ms) and negligible packet loss (0.05%).

The BPNN architecture exhibited robustness in handling complex data fusion tasks, achieving an impressive average quality of fused data at 98.6%. This underscores its capability to accurately integrate and process sensor data, critical for various applications in WSNs. In conclusion, the proposed approach offers a promising solution for enhancing data fusion in WSNs by effectively balancing energy efficiency and data accuracy. Future research directions may explore further optimization of the BPNN model parameters and its applicability across diverse WSN scenarios to harness its full potential.

## References

Akyildiz, I. F., Su, W., Sankarasubramaniam, Y., & Cayirci, E. (2002). Wireless sensor networks: A survey. Computer Networks, 38(4), 393-422.

Al-Karaki, J. N., & Kamal, A. E. (2004). Routing techniques in wireless sensor networks: A survey. IEEE Wireless Communications, 11(6), 6-28.

Biswas, S., & Misra, S. (2021). Advanced Machine Learning Algorithms for Energy Efficiency in WSN. IEEE Transactions on Green Communications and Networking, 5(2), 615-624.

Heinzelman, W. R., Chandrakasan, A., & Balakrishnan, H. (2000). Energy-efficient communication protocol for wireless microsensor networks. In Proceedings of the 33rd annual Hawaii international conference on system sciences (pp. 10-pp). IEEE.

Intanagonwiwat, C., Govindan, R., & Estrin, D. (2000). Directed diffusion: A scalable and robust communication paradigm for sensor networks. In Proceedings of the 6th annual international conference on Mobile computing and networking (pp. 56-67)

Liu, X., Xu, X., & Peng, Z. (2018). An energy-efficient clustering algorithm in wireless sensor networks with edge computing. IEEE Access, 6, 26145-26158.

Mohammadi, F., Taheri, H., & Yari, M. (2021). A survey on data aggregation techniques in IoT sensor networks. Journal of Network and Computer Applications, 176, 102919.

Prim, R. C. (1957). Shortest connection networks and some generalizations. Bell System Technical Journal, 36(6), 1389-1401.

Raza, U., Kulkarni, P., & Sooriyabandara, M. (2017). Low power wide area networks: An overview. IEEE Communications Surveys & Tutorials, 19(2), 855-873.

Rumelhart, D. E., Hinton, G. E., & Williams, R. J. (1986). Learning representations by back-propagating errors. Nature, 323(6088), 533-536.

Shafiq, M., Gu, Z., & Wang, H. (2019). Network traffic classification techniques and comparative analysis using machine learning algorithms. In Proceedings of the 2019 IEEE International Conference on Big Data (Big Data) (pp. 5717-5720).

Sun, Y., Ma, Y., & Wang, Y. (2022). Data fusion in wireless sensor networks: A survey. Information Fusion, 76, 38-50

Tahir, M., Kakkasageri, M. S., & Manvi, S. S. (2020). Minimum spanning tree-based cluster head selection for lifetime enhancement in wireless sensor networks. IEEE Sensors Journal, 20(5), 2751-2760.

Wang, J., Cao, Y., Xie, Y., & Wen, J. (2021). An Improved Fuzzy Clustering Algorithm for Wireless Sensor Networks. Sensors, 21(14), 4869.

Xu, N., Rangwala, S., Chintalapudi, K. K., Ganesan, D., Broad, A., Govindan, R., & Estrin, D. (2004). A wireless sensor network for structural monitoring. In Proceedings of the 2nd international conference on Embedded networked sensor systems (pp. 13-24).

Yang, Y. (2010). Fuzzy logic applications in wireless sensor networks. In Proceedings of the International Conference on Information and Automation (pp. 140-144). IEEE.

Younis, O., & Fahmy, S. (2004). HEED: A hybrid, energy-efficient, distributed clustering approach for ad hoc sensor networks. IEEE Transactions on Mobile Computing, 3(4), 366-379.